\documentclass[Times New Roman,10pt,a4paper,twocolumn]{article}
\usepackage[dvips]{graphicx}
\usepackage{amssymb,amsmath,epsfig}
\usepackage{url}

\usepackage{amsmath}
\usepackage{amssymb}
\usepackage{amsfonts}
\usepackage{graphicx}
\usepackage{verbatim}
\usepackage{algorithmic}
\usepackage{algorithm}
\usepackage{url}
\usepackage{longtable}

\def\argmax{\operatornamewithlimits{arg\,max}}

\title{Exploiting the Accumulated Evidence for Gene Selection
in Microarray Gene Expression Data}

\author{G. Prat\footnote{A shorter version of this paper appeared in the Procs. of the 19th European Conference on Artificial Intelligence (ECAI 2010).}, Ll. Belanche\footnote{Corresponding author.}\\
  School of Computer Science\\Polytechnical University of Catalonia\\Barcelona, Spain\\
  {\small\tt \{gprat, belanche\}@lsi.upc.edu} }

\date{}

\begin{document}

\maketitle

\begin{abstract}
  Machine Learning methods have of late made significant efforts to
  solving multidisciplinary problems in the field of cancer
  classification using microarray gene expression data. Feature subset
  selection methods can play an important role in the modeling
  process, since these tasks are characterized by a large number of
  features and a few observations, making the modeling a non-trivial
  undertaking.  

  In this particular scenario, it is extremely important to select
  genes by taking into account the possible interactions with other
  gene subsets. This paper shows that, by accumulating the evidence in
  favour (or against) each gene along the search process, the obtained
  gene subsets may constitute better solutions, either in terms of
  predictive accuracy or gene size, or in both. The proposed technique
  is extremely simple and applicable at a negligible overhead in cost.
\end{abstract}

\section{Introduction}
In the last years research in \emph{feature subset selection} (FSS)
has become a hot topic, boosted by the introduction of new application
domains and the growth of the number of features involved
\cite{Liu98}. An example of these new domains is web page
categorization, a domain currently of much interest for internet
search engines where thousands of terms can be found in a
document. Another example is found in cancer classification by gene
expression using DNA microarrays, a domain where Machine Learning
methods are now extensively used for this task \cite{Bo05}. Problems
with many features and a limited number of observations are also very
common in molecule classification or medical diagnosis, among others.

The selection of a new feature (either to be removed or added to the
current set) involves the evaluation of many models. These models
typically consist of the addition (deletion) of one feature to (from)
the current set. In \emph{wrapper} methods, an inducer is called to
build temporary solutions and return their evaluation using some
resampling method (e.g. cross-validation) \cite{KohaviJ97}. 

In the standard procedure, only the \emph{best} such model evaluation
is considered for selecting which feature should removed or added, and
the remaining evaluations are readily \emph{discarded}. Yet there
\emph{is} valuable information in the discarded evaluations: the very
many evaluated subsets contain information on the relevance of the
features that belong to the subset; this relevance does not depend on
the subset being selected or not. When an inducer is requested to
estimate the predictive accuracy of a model using a given feature
subset within a wrapper strategy, no indication is given on which
feature is the most recent addition (or deletion): the inducer just
sees a feature subset which has to be evaluated \emph{as a
  whole}. 

Since the most difficult part of a FSS process is to
evaluate the \emph{interactions} between features, the
\emph{accumulated} evaluation of a feature in diverse contexts should
account for many of these interactions, and ultimately provide with a
more informed estimation of usefulness for the chosen inducer. The
different \emph{contexts} of a particular feature $x$ are given by all
those subsets which are being evaluated along the search process (not
necessarily to assess the influence of $x$, as noted above), either
containing or \emph{not} containing $x$.

Our idea is to accumulate the inducer evaluations as a rich source of
information. This information can then be used in conventional
existing algorithms, such as the well-known \emph{forward} or
\emph{backward} selection. This idea can be applied to any sequential
search algorithm and any inducer and, as shown below, at a negligible
extra cost.

In this paper we present experimental results showing good performance
in a suite of benchmark microarray problems. The proposed modification
always achieves improvements when applied to standard backward
selection, either in the estimated predictive accuracy, in the size of
the delivered gene subsets, or in both.

\section{Accumulated Evidence in Feature Subset Selection}
\subsection{Preliminaries}

It is common to see feature subset selection (FSS) in a set $Y$ of
size $n$ as an {\em search problem} where the search space is the
power set of $Y$, $\mathcal P(Y)$ \cite{langley94selection}. Each
state in the search space corresponds to a subset of features.
Exhaustive search is usually intractable, and methods to explore the
search space efficiently must be employed.  These methods are often
divided into two main categories: \emph{filter} methods and
\emph{subset selection} methods. A major disadvantage of filter
methods is that they are performed independently of the classifier,
and the same set of features need not be optimal for different
classifiers.  Most filter methods disregard the dependencies between
features, as each feature is considered in isolation.

Without loss of generality, it can be assumed that the evaluation
measure $J:{\cal P}(Y) \rightarrow \mathbb{R}^{+} \cup \{ 0 \} $ is to
be maximized. In this setting, the problem is to find the optimal
subset $X^∗ \in P(Y )$ as the one maximizing $J$. The evaluation
measure maybe inducer-independent (as in filter methods) or may be the
same inducer being used to solve the task (as in wrapper methods). In
either case, we will refer to $J_{\mathcal L}(X)$ as the usefulness of
$X \subseteq Y$ estimated using the inducer ${\mathcal L}$. Since the
inducer evaluation in a sample varies depending on the resampling
method used, we prefer to use the notation $J_{\mathcal L}(X)$ instead
of simply ${\mathcal L}(X)$ to express such evaluation.

In the literature, several suboptimal algorithms have been proposed
for doing this. Among them, a wide family is formed by those
algorithms which, departing from an initial solution, iteratively add
or delete features by locally optimizing the objective function.  The
search starts with an arbitrary set of features (e.g. the full set or
the empty set) and moves iteratively to neighbor solutions by adding
or removing features. Among the most used algorithms for this problem
are the \emph{sequential forward generation} (SFG) and
\emph{sequential backward generation} (SBG), their generalization
\emph{plus $l$ - take away $r$} or $PTA(l,r)$
\cite{conf/icpr/Stearns76} \footnote{SFG is PTA(1,0) and SBG is
  PTA(0,1).} or the \emph{floating search} methods
\cite{DBLP:journals/prl/PudilNK94}. These latter algorithms work by
combining SFG and SBG steps.

\subsection{Accumulated evidence and feature relevance}

The idea consists on accumulating the \emph{evidence} in favor or against a
feature, taking into account its \emph{history} of evaluations alongside
different feature subsets. A further explanation can be to extract the most of
every subset evaluation, normally the most costly part of a FSS
process. 

Let $Y_x = \{X \in {\mathcal P}(Y) | x \in X\}$ be the set of all feature
subsets of the initial set that contain a certain feature $x$ (note
that $|Y_x|=2^{n-1}$ for all $x \in Y$). 

Let $\mathcal L^+_x$ and $\mathcal L^-_x$ be the average evaluation of
all subsets containing and \emph{not} containing $x$:

$${\mathcal L^+_x} = \frac{1}{2^{n-1}} \sum_{X \in Y_x} J_{\mathcal L}(X)$$

$${\mathcal L^-_x} = \frac{1}{2^{n-1}} \sum_{X \not\in Y_x} J_{\mathcal L}(X)$$

Given an inducer ${\mathcal L}$ (either filter or wrapper) define, for
a given feature $x \in Y$, the \emph{relevance} of $x$ as:

\begin{equation} 
R_{\mathcal L} (x) = {\mathcal L}_x^+ - {\mathcal L}_x^-
\label{fullL}
\end{equation}

The above definition can be more compactly expressed as:

\begin{equation}
R_{\mathcal L} (x) = \frac{1}{2^{n-1}} \sum\limits_{X \not\in Y_x}
\left\lgroup J_{\mathcal L}(X \cup \{x\}) - J_{\mathcal L}(X)\right\rgroup
\label{fullLequiv}
\end{equation}

\textbf{Remark 1}. Defining feature relevance with expression
(\ref{fullLequiv}) is very attractive, since it captures feature
interactions in all possible ways. We take the freedom of presenting
an informal but hopefully illustrative analogy of what this measure
captures. Imagine we are willing to evaluate the average influence of
a \emph{basketball} player on a team scoring: we can compute the
difference in points that the team scores \emph{with} and
\emph{without} this player, no matter what other players are playing
in the player’s team. If this difference is positive, then we can
conclude that this player’s accomplishments are positive for the team;
otherwise we conclude that we should better sell the player at the
best possible price! Note that in this example, only subsets $X$ of
size 4 are considered and $Y \setminus X$ is the
bench\footnote{Incidentally, this way of ranking players (together
  with rebounds, assists, etc) is used in the NBA.}.

\textbf{Remark 2}. Full evaluation of expression (\ref{fullLequiv})
has an exponential cost in $n$, making it unfeasible for most
practical applications; an estimation is therefore mandatory via Monte
Carlo techniques, generating feature subsets randomly from a precise
probability distribution determined by the FSS algorithm being
used. Oddly, although $R_{\mathcal L} (x)$ takes into account all
possible feature interactions, by its very nature it does not capture
redundancy: two identical features will have the same relevance. This
is true even by making $J_{\mathcal L}$ cope with redundancy.
However, since a search algorithm will impose an order on the
evaluated feature subsets, the current state can be used to ascertain
redundancy, as will be shown below.

The above expressions can be conveniently generalized by considering a
\emph{weighing} function $w$:

\begin{equation}
R_{\mathcal L}^w (x) = \frac{\sum\limits_{X \not\in Y_x}
\left\lgroup J_{\mathcal L}(X \cup \{x\}) - J_{\mathcal L}(x)\right\rgroup w_x(X)}
{\sum\limits_{X \not\in Y_x} w_x(X)}
\label{fullLW}
\end{equation}

For example, the choice $w_x(X) = |X|/|Y| = |X|/n$ gives more
importance to improvements in $J_{\mathcal L}$ achieved in a scenario
with already many features (improving performance in such a case has a
certain merit); alternatively, one could choose $w_x(X) = J_{\mathcal
  L} (X)$; this choice expresses the belief that an improved
performance when $J_{\mathcal L} (X)$ is already high should be
rewarded, and less so when it is low (it has a much lower merit). Many
alternatives are possible and the best one (if such choice exists at
all) is at the moment an open question.  Note that eq. (\ref{fullLW})
reduces to eq. (\ref{fullL}) when $w_x(X) = 1$ for all $x$.

In the following, we present a practical method to approximate
this measure of relevance and integrate it in a SBG search algorithm
at no additional cost. The idea consists on accumulating the \emph{evidence}
in favor or against a feature by taking into account the \emph{history} of
evaluations throughout the search process.

\subsection{Practical computation of the accumulated evidence}

Let $X_k$ denote the current set, where $|X_k| = k$, for notational simplicity
(thus $X_0 = \emptyset$ and $X_n = Y$); let $X_{n-k}$ be the set of features
not in $X_k$, i.e. $X_{n-k} = Y \setminus X_k$. Assume first we are in front of
performing a \emph{forward} step. Given $X_k$, in a classical SFG, the set

\begin{equation}
  \Big\{J_{\mathcal L}(X_k \cup \{x\}) \: | \: x \in X_{n-k} \Big\}
\textrm{ is computed}
\end{equation}

and the feature $x' = \argmax\limits_{x \in X_{n-k}}{J_{\mathcal L}(X_k
\cup \{x\})}$ is selected. However, all the remaining information:

\begin{equation}
  \Big\{J_{\mathcal L}(X_k \cup \{x\}) \: | \: x \in X_{n-k}, x \neq x'
\Big\} \textrm{ is discarded,}
\end{equation}

yet sometime in the future these individual features $x$ (and
eventually $x'$ itself) will be considered again for inclusion or
exclusion from the current set in forward or backward steps,
respectively.

Conversely, in a \emph{backward step} the search algorithm is going to
evaluate a feature $x$ for possible exclusion from $X_{n-k}$ in such a
way that the set

\begin{equation}
  \Big\{J_{\mathcal L}(X_{n-k} \setminus \{x\}) \: | \: x \in X_{n-k} \Big\}
\textrm{ is computed}
\end{equation}

and the feature $x' = \argmax\limits_{x \in X_{n-k}} J_{\mathcal L}(X_{n-k}
\setminus \{x\})$ is selected for removal. Again, the information:

\begin{equation}
  \Big\{J_{\mathcal L}(X_{n-k} \setminus \{x\}) \: | \: x \in X_{n-k}, x \neq x'
\Big\} \textrm{ is discarded.}
\end{equation}

Yet, sometime in the future these individual features $x$ (and eventually
$x'$ itself) will be considered again for inclusion or exclusion
from the current set in forward or backward steps, respectively. Reasoning
in more general terms, the search algorithm always evaluates
a feature $x$ for possible inclusion in (or exclusion from) the current
subset using information about $x$.

Now let $P_{\mathcal L}$ denote the set of feature subsets that the
search algorithm has evaluated so far (implying a call to ${\mathcal
  L}$). Let $P_{{\mathcal L}|x} = \{X \in P_{\mathcal L} | x \in
X\}$. For every $x \in Y$, define the accumulated evaluations (or simply
\emph{accumulators}) as the Monte Carlo estimations:

\begin{equation}
\hat{{\mathcal L}}^+_x = \frac{\sum\limits_{X \in P_{{\mathcal L}|x}}
J_{\mathcal L}(X)w_x(X)}
{\sum\limits_{X \in P_{{\mathcal L}|x}} w_x(X)}
\label{fullLW-estim+}
\end{equation}

\begin{equation}
\hat{{\mathcal L}}^-_x = \frac{\sum\limits_{X \not\in P_{{\mathcal L}|x}}
J_{\mathcal L}(X)w_x(X)}
{\sum\limits_{X \not\in P_{{\mathcal L}|x}} w_x(X)}
\label{fullLW-estim-}
\end{equation}

which are approximations to the weighted versions of ${\mathcal
  L}^+_x$ and ${\mathcal L}^-_x$, respectively. These two approximated
values depend on the search algorithm, which determines the strategy
to traverse the search space. Different FSS algorithms (such as SFG or
SBG) provide different traces of evaluated subsets at any given number
of algorithmic steps. In these conditions, the impact of the
considered feature in the \emph{current subset} $X$ can be used to
ascertain redundancy and make it influence the search, by modullating
the effect of the accumulated evaluations. Consider now, for $\lambda
\in [0,1]$,

\begin{equation} 
\hat{R}^w_{\mathcal L} (x) = \frac{\lambda}{2}(\hat{\mathcal L}_x^+ - \hat{\mathcal
  L}_x^- +1) + (1 - \lambda)\hat{J}_{\mathcal L} (x),
\label{finalR}
\end{equation}

where $\hat{J}_{\mathcal L} (x) = J_{\mathcal L} (X \setminus \{x\})$
in a backward step (the effect of removing $x$ from $X$) and
$\hat{J}_{\mathcal L} (x) = J_{\mathcal L} (X \cup \{x\})$ in a
forward step (the effect of adding $x$ to $X$) and $\lambda$ is a free
parameter. This scheme generalizes conventional forward and backward
steps (as used by SFG, SBG or any other sequential algorithm) in two
ways:

\begin{enumerate}
\item By setting $\lambda=0$, the conventional forward and backward
  steps are recovered and both relevance and redundancy are evaluated
  using $\hat{J}_{\mathcal L} (x)$. By setting $\lambda=1$, a pure
  arithmetic average between $\hat{\mathcal L}_x^+$ and $1 -
  \hat{\mathcal L}_x^-$ is computed.

  For other values of $\lambda$, the \emph{search history} makes an
  influence on the search itself, conditioning the selection of
  features.  In this case, only a $1-\lambda$ fraction of the
  importance is assigned to the current subset evaluation.
\item The search history itself is formed by all known contexts in
  which the considered feature could appear or not (and not only by
  previous evaluations of the feature), thus conforming a broader
  picture of its true relevance.
\end{enumerate}

\textbf{Example}. Consider the following feature subset mask ($n=20$)
for a current feature subset $X_8 \subset Y$ where the $i$-th index is
$1$ when feature $x_i \in X_8$ and $0$ otherwise:

\begin{verbatim}
            10010010001010100101 
\end{verbatim}

signaling the presence of features number $1, 4, 7$, etc. An
evaluation $J_{\mathcal L} (X)$ of this subset is indeed expressing
how good is to have the first feature but not the second or the third,
also how good is to have the seventh feature but not the one before
the last, and so forth. For this reason, all the features in $Y$ (and
not only those in $X$) should have their accumulators updated every
time.

\section{A practical algorithm}

We illustrate the approach on the popular SBG search algorithm
(\textbf{Algorithm 1}) and give a practical implementation of the
previous ideas for it (SBG$^+$, \textbf{Algorithm 2}). In addition,
for simplicity of presentation, we fix $w_x(X) = 1$. In this case,
normalization simply amounts to a division by the number of performed
accumulations. The initialization of the accumulated relevances is 0
for all $x \in Y$. The results are first accumulated and then used;
for this reason, even in the first algorithmic step (the first
discarded feature) the behavior of both algorithms may start to
diverge. At the end of the FSS process, $n^+_x$ (resp. $n^-_x$) will
be the number of times that a feature subset (resp. not) containing
$x$ has been evaluated. Note that the computation is done at a
negligible overhead in cost; this is due to the fact that the inducer
is called \emph{exactly} the same number of times for SBG than for the
accumulated counterpart SBG$^+$.

\begin{algorithm}
\caption{SBG (inducer ${\mathcal L}$, feature set $Y$)} \label{alg:SBG}
\begin{algorithmic}[1]
\STATE $X_n \leftarrow Y$
\STATE $k \leftarrow 0$
\REPEAT
	\FORALL{$x \in X_{n-k}$}
           \STATE compute the set $\Big\{J_{\mathcal L}(X_{n-k} \setminus \{x\})\Big\}$
        \ENDFOR
	\STATE $x' \leftarrow \argmax\limits_{x \in X_{n-k}} J_{\mathcal L}(X_{n-k}
\setminus \{x\})$
	\STATE $X_{n-k} \leftarrow X_{n-k} \setminus \{x'\}$
        \STATE $k \leftarrow k + 1$
\UNTIL{$k=n$}
\STATE {\bf return} $\argmax\limits_{k = 1 \div n}{J_{\mathcal L}(X_k)}$
\end{algorithmic}
\end{algorithm}

\begin{algorithm}
  \caption{SBG$^+$ (inducer ${\mathcal L}$, feature set $Y$, $\lambda \in
    [0,1]$)} \label{alg:SBG+}
\begin{algorithmic}[1]
\STATE $X_n \leftarrow Y$
\STATE $k \leftarrow 0$
\STATE \{Initialize accumulators and counters\}
\STATE $\forall x \in Y, \hat{{\mathcal L}}^+_x \leftarrow
\hat{{\mathcal L}}^-_x \leftarrow 0$
\STATE $\forall x \in Y, n^+_x \leftarrow n^-_x \leftarrow 0$
\REPEAT
	\FORALL{$x \in X_{n-k}$}
          \STATE compute the set $\Big\{J_{\mathcal L}(X_{n-k} \setminus \{x\})\Big\}$
        \ENDFOR
        \STATE \{Update accumulators and counters\}
        \FORALL{$x \in Y$}
          \IF{$x \in X_{n-k}$} 
            \STATE $\hat{{\mathcal L}}^+_x  \leftarrow \hat{{\mathcal
                 L}}^+_x + \sum\limits_{y \in X_{n-k} \setminus \{x\}}
                 J_{\mathcal L}(X_{n-k} \setminus \{y\})$
            \STATE $n^+_x \leftarrow n^+_x + 1$
          \ELSE
            \STATE $\hat{{\mathcal L}}^-_x  \leftarrow \hat{{\mathcal
                   L}}^-_x + J_{\mathcal L}(X_{n-k} \setminus \{x\})$
            \STATE $n^-_x \leftarrow n^-_x + 1$
          \ENDIF
        \ENDFOR
	\STATE $x' \leftarrow \argmax\limits_{x \in X_{n-k}}
        \left\{\frac{\lambda}{2}(\hat{\mathcal L}_x^+/n^+_x -
          \hat{\mathcal L}_x^-/n^-_x +1)\right.$\\
        \STATE $\hspace*{2.5truecm} \left. + (1 - \lambda)\hat{J}_{\mathcal L} (X_{n-k} \setminus
          \{x\})\right\}$
	\STATE $X_{n-k} \leftarrow X_{n-k} \setminus \{x'\}$
        \STATE $k \leftarrow k + 1$
\UNTIL{$k=n$}
\STATE {\bf return} $\argmax\limits_{k = 1 \div n}{J_{\mathcal L}(X_k)}$
\end{algorithmic}
\end{algorithm}

\section{Experimental work}
Experimental work is now presented in order to assess the described
modifications using two sequential algorithms: SBG and its accumulated
counterpart SBG$^+$. The algorithms were implemented using the R
language for statistical computing \cite{R08}.

\section{Experimental settings}
Each full experiment consists of an \emph{outer} loop of
5x2-cross-validation (5x2cv) for model selection, as proposed by
several authors \cite{Dietterich1998Approximate,Alpaydin1999Combined}.
This procedure performs 5 repetitions of a 2-fold cross-validation.  It
keeps half of the examples out of the feature selection process and
uses them as a \emph{test set} to evaluate the final quality of the
selected features. For every fold and repetition of the outer
cross-validation loop, two feature selection processes are conducted
with the same examples, one with the original algorithm (SBG) and one
with the accumulated version (SBG$^+$).

Each feature selection iteration uses the 1-\emph{nearest-neighbor}
learner implementation in \cite{Venables02Modern} (which uses
Euclidean distance), \emph{linear discriminant analysis} (LDA) and the
\emph{Support Vector Machine} with radial kernel (SVM$_r$). The
parameters of the SVM (the regularization constant or cost and the
kernel width) are kept fixed to their default values in all the
experiments, since we are only interested in the influence that
different feature subsets have on the modelling\footnote{These values
  are 1 for the cost parameter and the inverse of the number of
  features for the smoothing parameter in the kernel.}.

The evaluation of these inducers is resampled in a second
(\emph{inner}) 5x2cv loop for a more informed estimation of
usefulness. In all cases, stratification is used to keep the same
proportion of class labels across the partitioned sets. After some
preliminary experiments, we set $\lambda=\frac{2}{3}$ in expression
(\ref{finalR}). It is very important to mention that there is no
stopping criterion in the algorithms: the two backward methods run
until all the features have been removed. Then the \emph{best} subset
in the obtained sequence of subsets is returned. This setting avoids
the specification of an a priori size for the solution. It also
eliminates the possibility that the accumulated algorithm performs
differently simply because it merely influences the stopping point.

Once the best feature subset is found (a different one in every outer
loop), this subset is evaluated in the corresponding test set. The final
test error (the one reported) is the mean of these 10 values.

\subsection{Benchmarking microarray data sets}

In a microarray gene expression context, there is a wide spectrum of
FSS algorithms. Commonly found methods fall into the \emph{filter} category:
a list of the top-ranked genes based on some inducer-free figure
of merit is generated, followed by and inductive process where a classifier
is incrementally evaluated \cite{Rui05}. This constitutes a fast and low
complexity approach. However, considering individual contributions
only can hinder the discovery of possible interactions between genes.

Many authors have claimed that the \emph{wrapper} approach, if
affordable, is preferable to the filter approach
(e.g. \cite{Liu98,KohaviJ97}). It is therefore of the greatest
importance to take the most of every evaluation of the inducer, which
is normally the more costly part.

Validation of the described approach uses five public-domain microarray
gene expression data sets, shortly described as follows:

\begin{enumerate}
\item \emph{Colon Tumor}: Used originally by \cite{Al99}, it consists
  of 62 samples of colon tissue, of which 40 are tumorous and 22
  normal, and contains 2,000 genes.
\item \emph{Leukemia}: Used first by \cite{Go99}, the training set
  consisted originally of 38 bone marrow examples (plus a further test
  set with 34 examples).  This set of examples has been merged to form
  a data sample of 72 examples, which are described by 7,129 probes:
  6,817 human genes and 312 control genes. The goal is to tell acute
  myeloid leukemia from acute lymphoblastic leukemia.
\item \emph{Lung Cancer}: Studied by \cite{Gor02}, the problem
  consists in distinguishing between malignant pleural mesothelioma
  and adenocarcinoma of the lung. There are 181 examples available,
  described by 12,533 genes.
\item \emph{Prostate Cancer}: This data set was used by \cite{Sig02}
  to analyze differences in pathological features of prostate cancer
  and to identify genes that might anticipate its clinical
  behavior. There are 181 examples and 12,600 genes.
\item \emph{Breast Cancer}: \cite{Ver02} studied 97 patients with
  primary invasive breast carcinoma; 24,481 genes were analyzed.
\end{enumerate}

These problems are hard for several reasons, in particular the
sparsity of the data, the high dimensionality of the feature (gene)
space, and the fact that very many features (the genes) are irrelevant
or redundant. In these situations, performing feature selection is at
best a delicate task that entails a very high risk of overfitting,
even when the full set features has been preprocessed to lower the
dimensionality of the problem.

We made a preliminary selection of genes on the basis of the ratio of
their between-groups to within-groups sum of squares, as in other
approaches, to make a wrapper approach computationally feasible
\cite{Dudoit02Comparison}. In this work, the top 200 genes for each
dataset were selected as the source of study. It is important to
stress that there has been little effort to find the best models among
those represented by the considered inducers: in other words,
nearest-neighbors is limited to just one neighbour and the SVM
parameters have been set to their default values. All the effort is
devoted to find good feature subsets and to compare the two search
algorithms in similar experimental circumstances.

For comparative purposes, performance results using the whole set of
features and the reduced subset of 200 features are displayed in Table
\ref{Table:T200}. In view of these results, it is clear that these
subsets constitute a very good departing point for further analysis
with wrapper methods.

\begin{table}[!htb]
  \centering
  \begin{footnotesize}
  \begin{tabular}{|@{~}r@{~}|r@{~}r|r@{~}r|r@{~}r|}
  \hline
    & \multicolumn{ 2}{|c|}{1NN} & \multicolumn{ 2}{|c|}{LDA} & \multicolumn{ 2}{|c|}{SVM$_r$} \\ 
  \cline{2-7}
  Problem & $Y$ &  $X_{200}$ &  $Y$ & $X_{200}$ & $Y$ & $X_{200}$ \\
  \hline\hline
  Colon Tumor     &   23.9 & 23.2 & 24.8 & 20.0 & 31.0 & 14.8 \\ \hline
  Leukemia        &    9.7 &  8.3 & 14.1 &  3.1 & 26.7 &  2.8 \\ \hline
  Lung Cancer     &    1.8 &  2.0 & N/A  &  1.8 &  4.4 &  1.0 \\ \hline
  Prostate Cancer &   23.4 & 19.1 & N/A  & 25.5 & 38.2 & 26.9 \\ \hline
  Breast Cancer   &   45.1 & 27.7 & N/A  & 24.5 & 48.3 & 24.1 \\ \hline 
  \hline
\end{tabular}
  \end{footnotesize}
  \caption{Average test error (in \%) for the different inducers in the
    preprocessing phase. $Y$: using the full set of genes; $X_{200}$: using
    the top pre-selected 200 genes; N/A: computation unaffordable due to
    numerical inaccuracies in LDA.}
\label{Table:T200}
\end{table}

\begin{table}[!htb]
  \centering
  \begin{footnotesize}
  \begin{tabular}{|@{}r@{}|r@{}r|r@{}r|r@{}r|}
  \hline
    & \multicolumn{ 2}{|c|}{1NN} & \multicolumn{ 2}{|c|}{LDA} & \multicolumn{ 2}{|c|}{SVM$_r$} \\ 
  \cline{2-7}
  Problem & SBG$^+$ & SBG & SBG$^+$ & SBG & SBG$^+$ & SBG \\
  \hline\hline
  Colon Tumor     &   18.1 & 20.0 & 19.0 & 22.2 & 18.1 & 18.7 \\ \hline
  Leukemia        &    8.1 & 10.9 & 16.7 & 17.7 & 7.8 & 9.2   \\ \hline
  Lung Cancer     &    3.3 & 3.4 & 2.7 & 3.4 & 3.4 & 3.5      \\ \hline
  Prostate Cancer &   14.0 & 15.5 & 24.8 & 26.4 & 21.9 & 22.0 \\ \hline
  Breast Cancer   &   26.2 & 29.3 & 27.4 & 36.7 & 23.7 & 25.6 \\ \hline
  \hline
  Average         &   13.9 & 15.8 & 18.1 & 21.3 & 15.0 & 15.8\\
  \hline
\end{tabular}
  \end{footnotesize}
  \caption{Average test error (in \%) for the different inducers when
comparing SBG$^+$ to SBG.}
\label{Table:testerror}
\end{table}

\begin{table}[!htb]
  \centering
  \begin{footnotesize}
  \begin{tabular}{|@{}r@{}|r@{}r|r@{}r|r@{}r|}
  \hline
    & \multicolumn{ 2}{|c|}{1NN} & \multicolumn{ 2}{|c|}{LDA} & \multicolumn{ 2}{|c|}{SVM$_r$} \\ 
  \cline{2-7}
  Problem & SBG$^+$ & SBG & SBG$^+$ & SBG & SBG$^+$ & SBG \\
  \hline\hline
  Colon Tumor     &   37.4 & 73.8 & 70.5 & 79.2 & 15.5 & 14.2 \\ \hline
  Leukemia        &    7.2 & 28.3 & 30.0 & 32.5 & 6.1 & 37.2  \\ \hline
  Lung Cancer     &   17.4 & 20.0 & 4.1 & 13.4 & 4.5 & 8.8    \\ \hline
  Prostate Cancer &   18.3 & 19.3 & 23.5 & 44.3 & 12.9 & 8.1  \\ \hline
  Breast Cancer   &   60.2 & 34.2 & 22.4 & 52.6 & 13.0 & 17.5 \\ \hline
  \hline
  Average         &   28.1 & 35.1 & 30.1 & 44.4 & 10.4 & 17.2\\
  \hline
\end{tabular}
  \end{footnotesize}
  \caption{Average gene subset sizes for the different inducers when
comparing SBG$^+$ to SBG.}
\label{Table:sizes}
\end{table}

\section{Discussion}

The results of the FSS process are displayed in Tables
\ref{Table:testerror} and \ref{Table:sizes}. The first table shows the
(cross-validated) average test error for the two algorithms and the
different inducers. The second table shows the (cross-validated)
average size of the final selected subsets.

The first fact to note is that the accumulated version outperforms the
standard version (though in general by a modest margin) in all
cases. This is a very remarkable result, given the big differences
among the problems and among the inducers. Second, SBG$^+$ finds in
general solutions of lower size than SBG does, sometimes by a
substantial amount (e.g., 1NN in \emph{Colon Tumor} and
\emph{Leukemia}, most of LDA, or \emph{Leukemia} and \emph{Lung
  Cancer} with the SVM). Given that there is no stopping condition,
our explanation is that the standard backward version is \emph{greedier} than
the accumulated one. By the (early) inclusion of some (or many)
features that are not as good as they look in that moment, and cannot
be removed, SBG is driven toward worse local minima of the error
function as compared to SBG$^+$. The greediness itself is explained by
the purely \emph{local} (in the temporal sense) character of SBG and it also
explains the worse prediction results of this algorithm.


Feature selection appears to be a viable avenue for dimensionality
reduction in this field: a reduction of two orders of magnitude in the
number of features by univariate methods shows substantial
improvements (Table \ref{Table:T200}). With a further reduction of
another order of magnitude, mean performance of the finally selected
classifiers is similar to that achieved using the previously reduced
subset. This behavior is important, both for computational and
scientific reasons.  Even without optimization of free parameters (a
necessary step in normal conditions), cross-validated wrapper
computations with 200 features may take several days of computing time
on a modest machine. Scientifically, coping with hundreds of features
and pretending interpretability of the role of every feature in the
model is out of the question in many cases. This is aggravated in the
present situation of data scarcity.

The results diverge for different classifiers, as it may be reasonably
expected. This is of the greatest importance when assessing whether an
improvement is consistent, or is limited to a certain type of
method. In this sense, 1NN seems to be the best method for
\emph{Prostate Cancer}, LDA for \emph{Lung Cancer} and the SVM for the
other three (in all cases using SBG$^+$). The SVM tends to deliver
smaller gene subsets, both for SBG and SBG$^+$. Given that the SVM
parameters were not optimized beyond educated guesses, we think there
is room for further improvement in the modeling, specially on the
accuracy side.

Comparison to other results in the literature using the same data sets
is a delicate undertaking in general. The methodological steps can be
very different, especially concerning resampling techniques. We have
found that many times there are no true test sets: feature subsets or
model parameters (or both) are optimized by means of one or several
resampled runs of cross-validation. This procedure is dangerous in
that it cannot deliver an unbiased estimation of true error, given
that, although test observations have not been used to create the
model, they have been used to decide upon competing ones (namely, in
the feature selection process itself). The stability of these results
is also compromised if only one resample is carried out. On the other
hand, the delivered gene subset size is a very important issue to bear
in mind, if the solutions are to become interpretable and useful from
the clinical point of view. That said, we compare with several
references illustrative of recent work:

\begin{enumerate}
  \item For the \emph{Colon Tumor} data set, \cite{Li08} report an
  error of 12.7\% with 94 genes, while \cite{Bu07} report an error of
  23.0\% with 33 genes, both using radial SVMs. For this dataset, we
  report a test error of 18.1\% using an average of 15 genes. 

\item For the \emph{Leukemia} problem, \cite{Bu07} report an error of
  4.0\% with 30 genes using a radial kernel, and an extraordinary
  1.4\% using only two genes and filter methods for ranking
  \cite{Rat08}. For this dataset, we report an average test error of
  6.1\% using an average of 6 genes.

\item The \emph{Lung Cancer} data set is apparently the easiest to
  separate. Accuracy values as high as 99\% are achieved by
  \cite{Bu07} (using a SVM and 38 genes) and by \cite{Jin08}, this
  time using 5NN and as much as 135 genes. For this dataset, we report
  an average test error of 2.7\% using an average of 4 genes.

\item In the \emph{Prostate Cancer} problem, as low as 7\% error as
  been reported (half our best result) using a radial SVM and 47 genes
  (nearly three times our result) \cite{Bu07}.

\item Finally, for the \emph{Breast Cancer} problem, an error of 21\%
  is reported using a radial SVM and 46 genes \cite{Bu07}, and an
  error of 32\% using again a SVM and 8 genes \cite{Rat08}. For this
  dataset, we report an average test error of 23.7\% using an average
  of 13 genes.
\end{enumerate}


\section{Conclusions}
This paper has presented a modification suitable for feature subset
selection algorithms that iteratively evaluate subsets of features, by
making them accumulate all the ``log of merit'' of the features in
quite different contexts. The idea consists in that the current subset
evaluation is not used directly to select the feature to add (or
remove), but to \emph{accumulate} information on the usefulness of the
feature in many contexts. The different contexts of a particular
feature $x$ are given by all those subsets that contain $x$ (they
express how good is to have $x$) and do not contain $x$ (they express
how good is not to have $x$). The accumulated information is then used
to decide which feature should be added or removed (namely, that
feature with the highest (lowest) accumulated usefulness which has not
yet been added (removed)).  Therefore, the search history makes an
influence on the search itself, conditioning the selection of
features. This view is consistent with the definition of a search
algorithm as a mapping from its history (including its present state)
to the set of possible moves. In these conditions, less importance is
assigned to the current subset evaluation than in a classical FSS
setting (where it is the \emph{only} source of information).  Our
experimental results indicate a general improvement in performance,
without any additional modelling effort.

Future work includes exploring SFG. The decision to study SBG in the
first place is consistent with the goal of discovering feature
interactions. Having all the features from the beginning greatly
facilitates this task. Nonetheless, the more modest computational
demands that SFG entails in practice (if cut before exhaustion of
features) may be an appealing characteristic. It is relevant to point
out that the presented algorithmic modification may be of little help
if an algorithm has many opportunities to rectify its decisions
(\emph{e.g.}, the PTA$(l, r)$ family of algorithms). However, even in
this case, the forward or backward steps will be more informed,
possibly making the search algorithm deliver better solutions at
earlier stages. Unfortunately, the $O(n^{l+r+1})$ cost of PTA$(l, r)$
can well make it prohibitively high for microarray data problems in
wrapper mode.

A clear avenue for further research is the setting of the free parameter,
$\lambda$. It is our conjecture that an adaptive value may deliver
better results. In this sense, the influence of past evaluations may be
different at early or last stages of a search process.

\bibliography{FSS}

\end{document}